\documentclass[prc,nofootinbib,twocolumn]{revtex4}
\usepackage{graphicx,dcolumn,array,bm}

\newcommand{\tfrac}[2]{{\textstyle{\frac{#1}{#2}}}}
\newcommand{\svec}[1]{\bm{#1}}
\newcommand{\ivec}[1]{\vec{#1}}

\begin{document}
\bibliographystyle{apsrev}


\title{Nuclear time-reversal violation and the Schiff moment of $^{225}$Ra}
\author{J. Dobaczewski}
\email[]{Jacek.Dobaczewski@fuw.edu.pl}
\affiliation{Department of Physics and Astronomy, University of Tennessee,
  Knoxville, TN 37996}
 \affiliation{Physics Division, Oak Ridge National Laboratory, P.O.~Box 2008,
  Oak Ridge, TN 37831}
 \affiliation{Joint Institute for Heavy-Ion Research, Oak Ridge, TN 37831}
 \affiliation{Institute of Theoretical Physics, Warsaw University,
  ul.~Ho\.{z}a 69, 00-681 Warsaw, Poland}
\author{J. Engel}
\email[]{engelj@physics.unc.edu}
\affiliation{Department of Physics and Astronomy, CB3255,
             University of North Carolina, Chapel Hill, NC  27599-3255}

\date{\today}
\begin{abstract}
We present a comprehensive mean-field calculation of the Schiff moment of the
nucleus $^{225}$Ra, the quantity which determines the static electric dipole
moment of the corresponding atom if time-reversal (T) invariance is violated in the
nucleus. The calculation breaks all possible intrinsic symmetries of the
nuclear mean field and includes, in particular, both exchange and direct
terms from the full finite-range T-violating nucleon-nucleon interaction, and
the effects of short-range correlations. The resulting Schiff moment,
which depends on three unknown T-violating
pion-nucleon coupling constants, is much larger than in $^{199}$Hg, the isotope
with the
best current experimental limit on its atomic electric-dipole moment.

\end{abstract}
%
%
\maketitle
%
%

The Standard Model of particle physics violates time-reversal (T)
invariance, but apparently only through a single phase in the
Cabibo-Kobayashi-Maskawa
matrix that mixes quark flavors.  The resulting T violation in
flavor-conserving observables is therefore very weak and static electric-dipole
moments (EDMs) of neutrons, electrons, or atoms, all of which are nonzero if T
is violated, have never been observed.  Standard-Model T violation is also too
weak to account for the baryon asymmetry of the universe, which must come from
as yet undiscovered physics.  Happily, most theories of what
lies beyond the Standard Model contain enough phases that flavor-conserving T
violation will be unsuppressed.  Current levels of sensitivity in EDM
experiments are already sufficient to rule out or pressure many extra-Standard
models, and it seems quite possible that with slightly improved sensitivity,
new T-violating physics will be discovered.

Some of the tightest constraints on T violation come from atomic EDM
experiments.  The best of these at present is an experiment \cite{romalis01}
with $^{199}$Hg, but  it has
become clear recently
\cite{spevak95,spevak97,engel99a} that	atoms with octupole-deformed nuclei are
potentially more sensitive than Hg.  The primary reason is that given any T
violation in the nucleon-nucleon interaction, an asymmetric nuclear
shape and an associated
parity doubling create a collective ``Schiff'' moment, a kind of
radially weighted dipole moment (see below).
[Earlier papers on
T violation in polar diatomic molecules \cite{sushkov78} and parity violation
in fission \cite{sushkov80}
were the first to consider the effects of asymmetric shapes on
fundamental symmetries. 
Reference \cite{haxton83} drew
early attention to the importance for EDMs of low-lying nuclear states with the same angular momentum as the
ground state but opposite parity.]
Because of screening by atomic
electrons, the Schiff moment, rather than the nuclear EDM, is the quantity that
directly induces an atomic EDM (at least in lowest order; see Ref.\
\cite{flambaum02}).  In nuclei with symmetric shapes, a collective contribution
to the Schiff moment develops only in fluctuations around that shape
\cite{engel99a,flambaum03}.

In this paper we calculate the Schiff moment of $^{225}$Ra, or more precisely
its dependence on any T-violating $\pi NN$ couplings, in a mean-field theory
that allows us to break all possible symmetries, consider a variety of
phenomenologically successful strong (Skyrme) interactions, implicitly include
the RPA polarization of the even-even core by the valence neutron, treat both
the direct and exchange parts of the full pion-mediated interaction responsible
for creating the Schiff moment, and include short-range two-body correlations
between nucleons that modify the effects of this T-violating interaction.
Though further refinements are possible, they will probably have to include
correlations beyond mean-field theory and/or careful and systematic work on
Skyrme functionals; the results presented here will not be easy to supersede.

Simpler calculations of Schiff moments have been attempted before. Ref.\
\cite{sushkov84} applied an independent-particle model in $^{199}$Hg,
$^{129}$Xe, and other symmetrically deformed or spherical isotopes.  Refs.\
\cite{dmitriev03,dmitriev04} carried out a much more sophisticated RPA-based calculation
in $^{199}$Hg; it's main drawback was the use of a single phenomenological
interaction that made it difficult to estimate uncertainty.  Ref.\
\cite{spevak97} made estimates in a particle-rotor model of the
enhancement due to octupole deformation, and in Ref.\ \cite{engel03} we applied
a preliminary version of our technique to $^{225}$Ra, an experiment on
which is in the works \cite{holt03}.  That paper, however,
assumed the range of the T-violating interaction to be zero, an especially
bad approximation for exchange matrix elements, and was unable
(obviously) to examine the effects of short-range $NN$ correlations.

We briefly review some definitions and ideas.  The Schiff moment is given by
\begin{equation}
\label{eq:def}
S \equiv \langle \Psi_0 | \hat{S}_z | \Psi_0 \rangle = \sum_{i \neq 0}
         \frac{\langle \Psi_0 | \hat{S}_z |\Psi_i \rangle
               \langle \Psi_i | \hat{V}_{PT} | \Psi_0 \rangle}
              {E_0 - E_i}
         + \text{c.c.}
,
\end{equation}
where $| \Psi_0 \rangle$ is the member of the ground-state multiplet with
$J_z=J=1/2$ (positive parity), the sum is over excited states, and $\hat{S}_z$ is the operator
\begin{equation}
\hat{S}_z
= \tfrac{e}{10} \sum_p \left( r_p^2 -
    \tfrac{5}{3} \overline{r_{\rm ch}^2}  \right) \, z_p
,
\end{equation}
with the sum here over protons, and  $\overline{r_{\rm ch}^2}$ the mean-square
charge radius. The operator $\hat{V}_{PT}$ in Eq.\ (\ref{eq:def}) is the T- (and
parity-\nolinebreak) violating nucleon-nucleon interaction mediated by the pion
\cite{haxton83,herczeg88}:
\begin{widetext}
\begin{eqnarray}
\label{eq:pion}
\hat{V}_{PT}(\svec{r}_1-\svec{r}_2)
               & = - {\displaystyle   \frac{g \, m_{\pi}^2}{8 \pi m_N}} &
                     \Big\{
                         (\svec{\sigma}_1 - \svec{\sigma}_2) \cdot
                         (\svec{r}_1-\svec{r}_2)
                         \left[ \bar{g}_0 \, \ivec{\tau}_1 \cdot
                                             \ivec{\tau}_2
                         - \frac{\bar{g}_1}{2} \,
                          (\tau_{1z}+\tau_{2z}) + \bar{g}_2
                         (3\tau_{1z}\tau_{2z} - \ivec{\tau}_1 \cdot
                         \ivec{\tau}_2 ) \right]
                            \\
                        && -\frac{\bar{g}_1}{2}
                         (\svec{\sigma}_1+\svec{\sigma}_2)
                         \cdot (\svec{r}_1-\svec{r}_2) \,
                         (\tau_{1z}-\tau_{2z})  \Big\}
\frac{{\rm exp}(-m_{\pi} |\svec{r}_1-\svec{r}_2|)}{m_{\pi}
|\svec{r}_1-\svec{r}_2|^2} \left[ 1+\frac{1}{m_{\pi}|\svec{r}_1-\svec{r}_2|}
\right] \nonumber
,
\end{eqnarray}
\end{widetext}
where arrows denote isovector operators, $\tau_z$ is +1 for neutrons,
$m_N$ is the nucleon mass, and (in this equation only) we
use the convention \mbox{$\hbar = c = 1$}.  The $\bar{g}$'s are the unknown
isoscalar, isovector, and isotensor T-violating pion-nucleon
coupling constants, and $g$ is the usual strong ${\pi NN}$ coupling constant.

\begin{figure}[t]
\includegraphics[width=5cm]{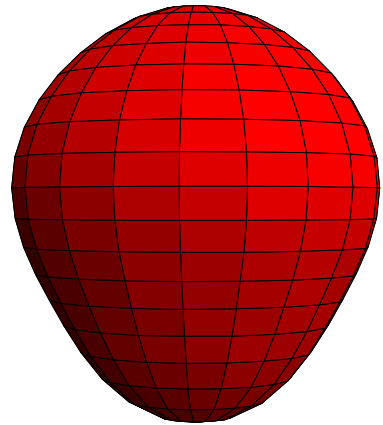}
\caption{\label{f:1}(color online).
Shape of the microscopically calculated \protect\cite{engel03} mass distribution
in $^{225}$Ra, represented here by the surface of a uniform body that
has the same multipole moments $Q_{\lambda0}$ for $\lambda$=0\ldots4 as our
calculated density.}
\end{figure}

The asymmetric shape of  $^{225}$Ra implies parity doubling
(see e.g.\ Ref.\ \cite{sheline89}), i.e.\ the existence of a very
low-energy $|1/2^-\rangle$ state, in this case 55\,keV \cite{helmer87} above the ground state
$|\Psi_0\rangle \equiv |1/2^+\rangle$, that dominates the sum in Eq.\
(\ref{eq:def}) because of the corresponding small denominator.  With the
approximation that
the shape deformation is rigid,
the
ground state and its negative-parity partner in octupole-deformed nucleus are
projections onto good parity and angular
momentum of the same ``intrinsic state" (see Fig. \ref{f:1}), which represents the wave function of
the nucleus in its own body-fixed frame
with the total angular momentum aligned along the symmetry axis.
  Equation (\ref{eq:def}) then reduces to \cite{spevak97}
\begin{equation}
\label{eq:intr} S \approx - \frac{2}{3} \langle
\hat{S}_z \rangle
\frac{\langle  \hat{V}_{PT} \rangle} { (55\,\mathrm{keV})}  \ \ ,
\end{equation}
where the brackets indicate expectation values in the intrinsic state.  Using
Eq.\ (\ref{eq:pion}) for $\hat{V}_{PT}$, we can express the dependence of the
Schiff moment on the undetermined T-violating $\pi NN$ vertices as
\begin{equation}
\label{eq:coefs}
S = a_0 \, g \, \bar{g}_0 + a_1 \, g \,  \bar{g}_1 +a_2 \, g \,  \bar{g}_2~.
\end{equation}
The coefficients $a_i$, which are the result of the calculation, have units $e$\,fm$^3$.

The octupole deformation enhances $\langle
\hat{S}_z \rangle$, the first factor in Eq.\ (\ref{eq:intr}), making it collective, robust, and straightforward to
calculate with an error of a factor of two or less.  The interaction
expectation value $\langle  \hat{V}_{PT} \rangle$ is harder to estimate because it is
sensitive to the nuclear spin distribution, which depends on delicate
correlations near the Fermi surface.  Our calculation allows the
breaking of Kramers degeneracy in the intrinsic frame and, consequently, spin
polarization.

\begin{figure}[b]
\includegraphics[width=8cm]{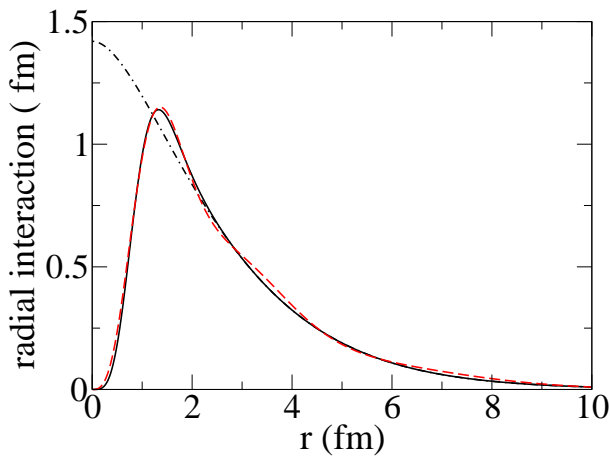}
\caption{\label{f:2}(color online).
The function $ g(r)$ in Eq.\ (\ref{eq:g}) multiplied by
$r^3$ (solid line), the
Gaussian fit multiplied by $r^3$ (dashed line), and $r^3 g(r)/f(r)^2$, the
radial T-odd interaction
without short-range correlations (dot-dashed line).  The factor $r^3$ is to account
for the volume element and the additional factor of $\svec{r} \equiv
\svec{r}_1-\svec{r}_2$ in Eq.\ (\ref{eq:pion}).}
\end{figure}

To evaluate $\langle  \hat{V}_{PT} \rangle$ we constructed a new version of
the code HFODD (v2.14e) \cite{dobaczewski04,hfodd}. The code
uses a triaxial harmonic-oscillator basis and Gaussian integration
to solve self-consistent mean-field equations for zero-range Skyrme interactions.
Evaluating matrix elements of the finite-range interaction (\ref{eq:pion}) is much harder numerically, but
efficient techniques have already been developed \cite{girod83} for Gaussian interactions,
which are separable in three Cartesian directions.  The spatial dependence in
Eq.\ (\ref{eq:pion}) is different, the derivative of a Yukawa function, and we
also
include short-range correlations between nucleons (which the mean-field does
not capture) by multiplying the interaction by the square of a correlation
function \cite{miller76} that cuts off the two-nucleon wave functions below a
relative distance of about a fermi:
\begin{equation}
\label{eq:src}
f(r) = 1-e^{- 1.1 r^2}(1- .68 \, r^2)~,
\end{equation}
with $r \equiv |\svec{r}_1 - \svec{r}_2|$ in fermis and the coefficients of $r^2$ in fm$^{-2}$.  The resulting
product looks very different from a Gaussian, but we were able to reproduce it
quite accurately (see Fig.\ \ref{f:2}) with the sum of four Gaussians:
\begin{widetext}
\begin{equation}
\label{eq:g}
g(r)= f(r)^2 \, \frac{e^{-a_{\pi} r}}{r^2} \left( 1 + \frac{1}{a_{\pi}
r} \right)
\approx  1.75\, e^{-1.1  r^2}
 + 0.53 \, e^{-0.68 r^2} \!\!
  +  0.11 \, e^{-0.21 r^2} \!\!
  + 0.004 \, e^{-0.06  r^2} \!,
\end{equation}
\end{widetext}
where $a_{\pi} \equiv
0.7045~\mathrm{fm}^{-1}$ is the pion mass in inverse fermis and
the numbers in the fit all have units fm$^{-2}$. The extra
factor of $\svec{r}$ not included in Eq.\ (\ref{eq:g}) (i.e.\ the factor
$\svec{r}_1-\svec{r}_2$ in Eq.\
(\ref{eq:pion})) is treated separately.

HFODD works with any Skyrme energy functional. In the context of the present study
the best is
SkO$^{\prime}$ \cite{reinhard99,bender02}.  The ``time-even'' terms in this
interaction, which act in even-nucleus ground states, were fit with special
attention to nuclei around $^{208}$Pb and to spin-orbit splitting.  The
``time-odd'' terms responsible for core-polarization in an odd nucleus
were adjusted in Ref.\ \cite{bender02} to
reproduce Gamow-Teller resonances, resulting in an effective Landau parameter
$g_0^{\prime}=1.2$. (The isoscalar parameter was set to $g_0=0.4$, following
common practice.)  For comparison, we also carry out the calculation with the
older parameterizations SIII, SkM*, and SLy4 (with time-odd terms that are
not fixed by gauge invariance neglected, and then again, with the simplest
time-odd terms modified so that the Landau parameters
have the same values as in SkO$^{\prime}$) but have the most confidence in
SkO$^{\prime}$.  Ref.\ \cite{engel03} presented predictions by these
functionals for the binding energies, separation energies, intrinsic dipole
moments, and spin-orbit splittings in the even Ra isotopes.  SkO$^{\prime}$ and SIII
seemed to do the best job.

\begin{table}[h,b]
\caption{\label{t:1} Coefficients of $g \bar{g}_i$, in units of
$e\,\textrm{fm}^{3}$, in the expression Eq.\ (\ref{eq:coefs}) for the Schiff
moment of $^{225}$Ra, calculated with the SkO$^{\prime}$ Skyrme interaction.
The abbreviation ``src'' stands for ``short-range correlations''.}
\begin{ruledtabular}
\begin{tabular}{lccc}
 & $a_0$ & $a_1$ & $a_2$\\
\hline
zero-range (direct only)   &  $-$5.1 &  10.4 &$-$10.1 \\
finite-range (direct only)&  $-$1.9 &  6.3 &$-$3.8 \\
finite-range + src (direct only)  &  $-$1.7 &  6.0 & $-$3.5 \\
\hline
finite-range + src (direct + exchange) &  $-$1.5 &  6.0 & $-$4.0 \\
\end{tabular}
\end{ruledtabular}
\end{table}

Table \ref{t:1} shows the calculated values, with SkO$^{\prime}$, of the three
coefficients $a_i$ at several levels of approximation. The finite range reduces
the direct matrix elements of the interaction (and the corresponding $a_i$)
from the zero-range limit \cite{engel03} significantly. The exchange terms are
reduced much more, so that they are always smaller than the direct terms.  The
effects of the short-range correlations, which also reduce the coefficients,
are relatively small as well but non-negligible. Finally, as pointed out in
Ref.\ \cite{flambaum02}, relativistic effects in electron wave functions
correct the effects of the Schiff moment; the authors summarize the corrections
in a quantity they call the ``local nuclear dipole moment''.  Our local dipole
moment in $^{225}$Ra is 82\% of the Schiff moment for the $s_{1/2}-p_{1/2}$
atomic transition and 87\% for the $s_{1/2}-p_{3/2}$ transition.

\begin{table}[t,h]
\caption{\label{t:2} Same as the last line in Table \protect\ref{t:1} but for the
SIII, SkM*, and SLy4 interactions.}
\begin{ruledtabular}
\begin{tabular}{lccc}
 & $a_0$ & $a_1$ & $a_2$\\
\hline
SIII  &  $-$1.0 &  7.0  &$-$3.9   \\
SkM*  &  $-$4.7 &  21.5 &$-$11.0  \\
SLy4  &  $-$3.0 &  16.9 &$-$8.8   \\
\end{tabular}
\end{ruledtabular}
\end{table}
The coefficients produced by
the forces we favor less, with the time-odd terms adjusted as mentioned
above, are presented in Tab.\ \ref{t:2}; SIII produces
numbers similar to those of SkO$^{\prime}$, while the other two forces give numbers
that are larger by factors of two or three, whether or not the Landau
parameters are adjusted (i.e., the effects of adjusting those parameters seem
to be fairly small).  

What is the uncertainty in our numbers?  The mean-field
omits correlations that could have some effect on the result; those could be
explored by refining the calculation through angular-momentum and parity
projection, i.e., the restoration of symmetries broken by the mean field.  In addition, an optimal Skyrme functional has yet to be
identified.  Those we tested give results that differ from the SkO$^{\prime}$
numbers by
factors two or three, as mentioned above.  But some low-order terms in
the T-odd part of the Skyrme functional are never used even in SkO$^{\prime}$,
because they have never been fit.  Ref.\ \cite{bender02} constrained some
combinations of those terms but others were set to zero for lack of sufficient
Gamow-Teller data in spherical nuclei.  One might imagine trying to fit in
deformed nuclei, or looking at spin-strength distributions with different total
angular momentum and parity; the $0^-$ channel would be particularly useful
because those are the quantum numbers of $\hat{V}_{PT}$.
At the same time, it would probably help to explicitly study the sensitivity of the
Schiff moments to changes in the various Skyrme parameters, both in the
time-odd and time-even sectors.  With enough work on all these fronts we could
give a firmer estimate of the uncertainty than our current
guess: a factor of two or three.

If we accept our current results as reasonably accurate, we are in a
position to quantify the advantages of $^{225}$Ra for an EDM
measurement.  A recent RPA calculation \cite{dmitriev03,dmitriev04} of the $a_i$ for $^{199}$Hg gives
\begin{equation}
{}^{199}\mathrm{Hg}: ~ ~ ~ a_0 = 0.0004~, ~ ~ ~  a_1 = 0.055~, ~  ~ ~
a_2 = 0.009.
\end{equation}
Our numbers are more than two orders of magnitude
larger, particularly in the isoscalar channel (the relevant channel if
T-violation is caused by a nonzero QCD parameter $\bar{\theta}$), where the Hg number is
anomalously small.
Atomic physics enhances any EDM in Ra by another
factor of 3 over that in Hg \cite{dzuba02}, so if the Ra EDM can be measured even 1/100th as
accurately as that of Hg, the sensitivity to nuclear $T$ violation will be
significantly greater.

In conclusion, we have evaluated the Schiff moment of $^{225}$Ra in a
completely symmetry-breaking mean-field approach, including in particular the
finite-range matrix elements of the T-violating nucleon-nucleon interaction.
The results indicate that EDM experiments in this system are very promising.
The remaining uncertainties of a factor of two or three are related primarily
to deficiencies in nuclear effective interactions, which can be removed but not
easily.

Fruitful discussions with J.~H.\ de Jesus and W.\ Nazarewicz are gratefully acknowledged.
This work was supported in part by the U.S.\ Department of Energy
under Contracts Nos.\ DE-FG02-96ER40963 (University of Tennessee),
DE-AC05-00OR22725 with UT-Battelle, LLC (Oak Ridge National
Laboratory), DE-FG05-87ER40361 (Joint Institute for Heavy Ion
Research), DE-FG02-97ER41019 (University of North Carolina); by the
Polish Committee for Scientific Research (KBN) under Contract
No.~1~P03B~059~27; and by the Foundation for Polish Science (FNP).

\end{document}